\documentclass[a4paper,11pt]{article}
\usepackage{pos}
\usepackage{graphicx}
\usepackage{amsmath}
\usepackage{hyperref}  

\newcommand{\dint}{{\rm d}}
\newcommand{\xpom}{x_{I\!\!P}}
\newcommand{\sartre}{Sar\emph{t}re}
\newcommand{\tsc}[1]{\textsc{#1}}
\newcommand{\pythia}{\tsc{Pythia}}

\title{Inclusive diffraction in eA at small x with the Sartre event generator}

\author*[a]{Tobias Toll}
\author[a]{Bhakta Naik}
\author[a]{Jaswant Singh}
\author[b]{Thomas Ullrich}

\affiliation[a]{Department of Physics, Indian Institute of Technology Delhi,\\
Hauz Khas, New Delhi 110 016, India}
\affiliation[b]{Brookhaven National Laboratory,\\ Upton, NY 11973, USA}

\emailAdd{tobiastoll@iitd.ac.in}
\emailAdd{thomas.ullrich@bnl.gov}

\abstract{%
We present an extension of the \sartre\ event generator that implements inclusive diffractive deep inelastic scattering for both proton and nuclear targets at small $x$.  The calculation is performed in the colour-dipole picture using the IPsat and IPnonsat dipole amplitudes that \sartre\ also employs for exclusive processes, and includes the transverse and longitudinal $q\bar q$ contributions together with the $q\bar qg$ Fock state that dominates at large $M_X$. The partonic diffractive system is hadronised with \pythia, giving fully exclusive hadronic final states.  We compare to the combined H1 and ZEUS inclusive diffractive measurements and to diffractive $D^*$ production, and present predictions for the Electron-Ion Collider.  This makes \sartre\ 2 the only event generator providing inclusive diffractive DIS in small-$x$ for both $ep$ and $eA$ collisions.}

\FullConference{%
}

\begin{document}
\maketitle

\section{Introduction}
A central question in high-energy QCD is the behaviour of matter at high gluon density.  HERA established that at small Bjorken $x$ the gluon density rises rapidly~\cite{H1:2009pze}; this growth must eventually be tamed by non-linear recombination, leading to gluon saturation~\cite{Iancu:2003xm,Weigert:2005us}. The Electron-Ion Collider (EIC)~\cite{Accardi:2012qut,AbdulKhalek:2021gbh} is designed to establish where this regime sets in, with nuclear beams for which the saturation scale grows roughly as $A^{1/3}$.

Diffraction is especially sensitive to saturation.  About ten percent of DIS events at HERA are diffractive, with a colour-neutral exchange and a large rapidity gap~\cite{H1ZEUS:2012diff}.  At small $x$ the diffractive cross section is approximately quadratic in the gluon density, against linear for inclusive DIS, making it one of the cleanest probes of the onset of saturation~\cite{Kowalski:2008sa}.  The colour-dipole picture describes inclusive DIS, exclusive diffraction and inclusive diffraction with one universal dipole amplitude, the processes differing only in the projection of the photon wave function after the interaction. Impact-parameter dependent versions such as IPsat extend naturally to nuclei and allow saturated and linearised dynamics to be compared in a common framework~\cite{Kowalski:2003hm,Kowalski:2006hc}.

The \sartre\ event generator~\cite{Toll:2012mb,Toll:2013gda} implemented this framework for exclusive vector-meson production and DVCS in $ep$ and $eA$, including coherent and incoherent nuclear diffraction, and was used extensively in developing the EIC physics and detector case.  Inclusive diffractive DIS, $e+p/A\rightarrow e'+p'/A'+X$, carries considerably more information: the photon dissociates into a system $X$ of mass $M_X$,  described by $\beta=Q^2/(Q^2+M_X^2)$, and at large $M_X$ the $q\bar qg$ Fock state becomes essential~\cite{Wusthoff:1999cr,Kowalski:2008sa}. Here we present \sartre\ 2, which implements inclusive diffractive DIS for proton and nuclear targets, including the transverse and longitudinal $q\bar q$ and the $q\bar qg$ contributions, with the final state hadronised by \pythia.

\section{The Dipole Model}
\label{sec:dipolemodel}
\begin{figure}[htbp]
    \centering
    \includegraphics[width=0.8\textwidth]{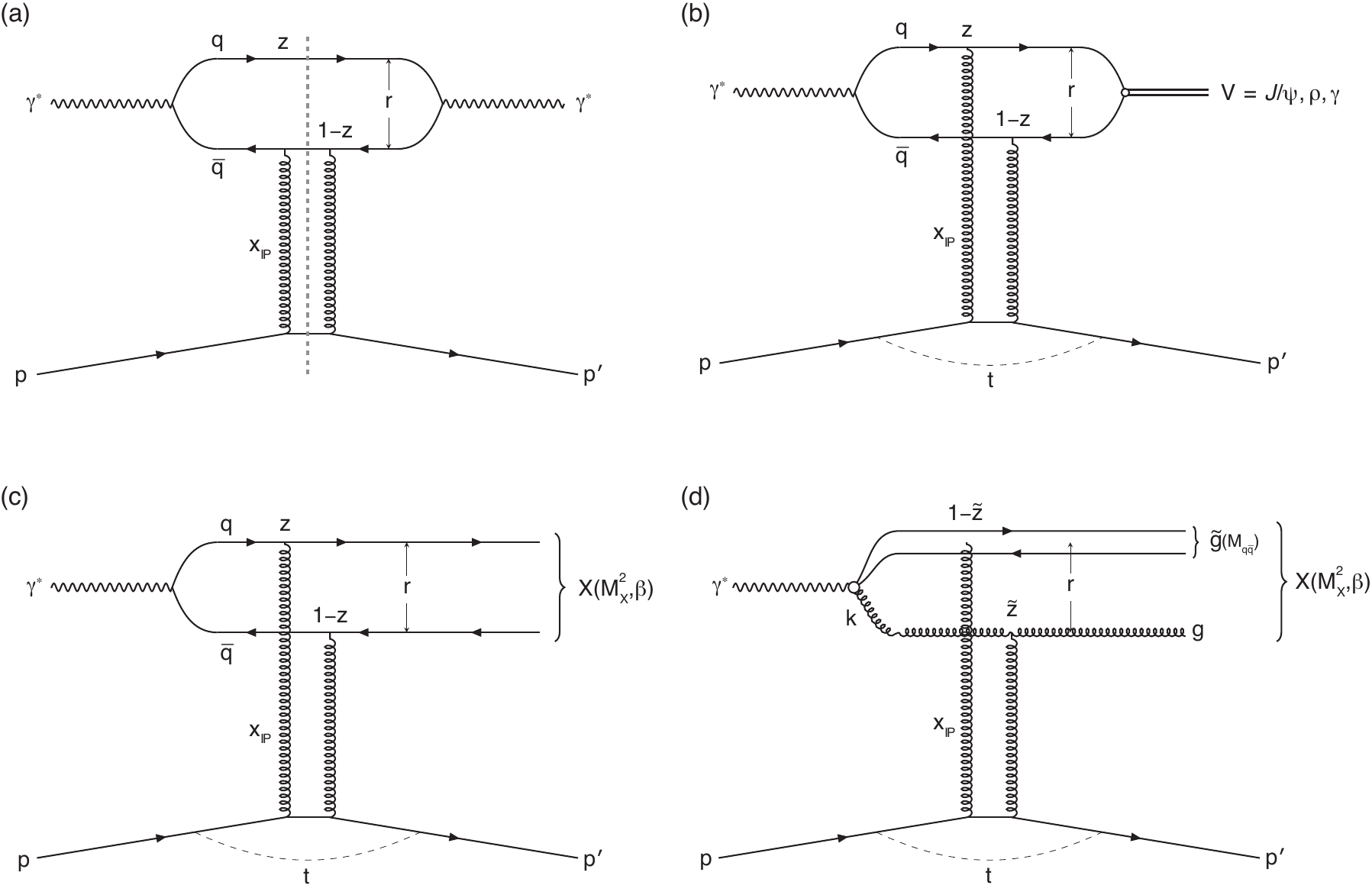}
    \caption{A pictorial view of the dipole model.  (a) Inclusive DIS,
    (b) exclusive diffraction, (c) inclusive diffraction, $q\bar q$ Fock state,
    (d) inclusive diffraction, $q\bar qg$ Fock state.}
    \label{fig:dipolemodels}
\end{figure}

Fig.~\ref{fig:dipolemodels} shows different processes in Deep-Inelastic Scattering (DIS). In inclusive DIS Fig.~\ref{fig:dipolemodels}(a) show the absolute square of the amplitude which can be depicted as the photon splitting into a $q\bar q$ dipole which after the interaction with the target recombined back into a virtual photon. Fig.~\ref{fig:dipolemodels}(b) depict the amplitude of exclusive diffraction where the dipole recombines into a vector meson or real photon. In inclusive diffraction (c), it instead forms a hadronic final state $X$, while (d) depicts the higher Fock-state where the photon splits into a $q\bar q g$ dipole before the interaction.  In diffraction the dipole exchanges a colour-neutral two-gluon state carrying momentum fraction $\xpom$, leaving the proton intact across a rapidity gap $\Delta y_{\rm gap}=\ln(1/\xpom)$.  The diffractive system spans $\Delta y_X=\ln(1/\beta)$, with $x=\beta\xpom$ and $M_X^2=Q^2(1-\beta)/\beta$.

In all cases the dipole--target interaction is governed by the same universal dipole amplitude $\mathcal{N}(r,x,\vec b)=\tfrac12\,\dint\sigma_{q\bar q}/\dint^2\vec b$. In \sartre\ we implement two versions,
\begin{equation}
    \frac{\dint\sigma_{q\bar q}^{\rm sat}}{\dint^2\vec b}(r, x, \vec b)
    =2\left[1-e^{-\Omega(r, x, \vec b)/2}\right],
    \qquad
    \frac{\dint\sigma_{q\bar q}^{\rm nosat}}{\dint^2\vec b}(r, x, \vec b)
    =\Omega(r, x, \vec b),
\end{equation}
where the opacity is
\begin{equation}
    \Omega(r,x,\vec b)=\frac{\pi^2}{N_C}\,r^2\,\alpha_S(\mu^2)\,x g(x,\mu^2)\,T(\vec b).
\end{equation}
Here $T(\vec b)$ is the target thickness function, i.e.\ the spatial gluon density projected onto the transverse plane.  The scale is $\mu^2=C/r^2+\mu_0^2$, and $xg$ is evolved with gluon-only DGLAP from an initial distribution $xg(x,\mu_0^2)=A_gx^{-\lambda_g}(1-x)^6$.  IPsat saturates at large $r$, large gluon density and large thickness; IPnonsat is its leading-twist expansion and does not saturate.  Comparing the two allows the amount of non-linearity in different regions of phase space to be assessed directly.

Both models have been fitted to the inclusive DIS reduced cross section at HERA~\cite{Rezaeian:2012ji,Mantysaari:2018nng,Sambasivam:2019gdd} and describe exclusive diffraction well~\cite{Kowalski:2006hc,Toll:2026hmx}, supporting the universality of the dipole cross section.  In inclusive diffraction, however, the $q\bar q$ dipole becomes subleading in parts of phase space and the higher Fock state $q\bar qg$ is needed, which constitutes a colour quadrupole, or equivalently an adjoint-representation dipole.  We describe both contributions below, following~\cite{Kowalski:2008sa}.

\subsection{The $q\bar q$ Fock state}
Define the amplitude
\begin{equation}
    \mathcal{A}_n=\int_0^\infty r\,\dint r\; K_n(\epsilon r)\,J_n(kr)\,
    \frac{\dint\sigma_{q\bar q}}{\dint^2\vec b},
\end{equation}
where $\epsilon^2=z(1-z)Q^2+m_f^2$ sets the size of the incoming dipole and $k^2=z(1-z)M_X^2-m_f^2$ is the transverse momentum of the produced quark, so that $K_n(\epsilon r)J_n(kr)$ is the overlap between the incoming photon and the outgoing $q\bar q$ pair.  With $\Phi_n=\int\dint^2\vec{b}\,|\mathcal{A}_n|^2$, the $\gamma^*p$ cross sections for transversely and longitudinally polarised photons are
\begin{align}
    \frac{\dint^2\sigma_T^{\gamma^*p}}{\dint\beta\,\dint z}&=
    \frac{N_C Q^2\alpha_{\rm EM}}{8\pi\beta^2}\sum_f e_f^2\, z(1-z)
    \left[\epsilon^2\big(z^2+(1-z)^2\big)\Phi_1+m_f^2\Phi_0\right],\\
    \frac{\dint^2\sigma_L^{\gamma^*p}}{\dint\beta\,\dint z}&=
    \frac{N_C Q^4\alpha_{\rm EM}}{2\pi\beta^2}\sum_f e_f^2\, z^3(1-z)^3\,\Phi_0,
\end{align}
where $z$ is the light-cone momentum fraction of the photon's $p^-$ taken by the quark, with $z_0\leq z\leq 1-z_0$ and $z_0=\tfrac12\big(1-\sqrt{1-4m_f^2/M_X^2}\big)$.  The $ep$ cross sections is achieved by multiplying by the standard transverse and longitudinal photon fluxes.

\subsection{The $q\bar qg$ Fock state}
At large $M_X^2$ the $q\bar qg$ Fock state becomes important, giving large logarithms in $Q^2$, in $1/\beta$, or in both.  In the large-$Q^2$ limit the quark and antiquark are close in impact-parameter space and can be treated as a pseudo-gluon $\tilde g$, giving an effective $\tilde gg$ colour quadrupole. The amplitude is
\begin{equation}
	\mathcal{A}_{\tilde gg}=\int r\,\dint r\; K_2\big(\sqrt{\tilde z}\,kr\big)
	J_2\big(\sqrt{1-\tilde z}\,kr\big)
	\frac{\dint\tilde\sigma_{\tilde gg}}{\dint^2\vec b},
\end{equation}
with the adjoint-representation dipole cross section
\begin{equation}
\frac{\dint\tilde\sigma_{\tilde gg}}{\dint^2\vec b}
=2\left[1-\left(1-\frac{1}{2}\frac{\dint\sigma_{q\bar q}}{\dint^2\vec b}\right)^2\right].
\end{equation}
Keeping the notation of the $q\bar q$ Fock state,
\begin{equation}
	\tilde \Phi_{\tilde gg}=\int_0^{Q^2}\dint k^2\, k^4 \ln\frac{Q^2}{k^2}
	\int \dint^2\vec{b}\,|\mathcal{A}_{\tilde gg}|^2,
\end{equation}
and the contribution to the cross section is
\begin{equation}
    \frac{\dint^2\sigma_{\tilde gg,T}^{\gamma^*p}}{\dint\beta\,\dint \tilde z}=
    \frac{\alpha_{\rm S}\alpha_{\rm EM}}{2\pi^2Q^2}
    \left[\left(1-\frac{\beta}{\tilde z}\right)^2+\left(\frac{\beta}{\tilde z}\right)^2\right]
    \sum_f e_f^2\,\tilde\Phi_{\tilde gg}.
\end{equation}
Here $\tilde z$ is the light-cone momentum fraction $p^+$ of the target taken by the gluon, $\beta\leq \tilde z\leq 1$, and $k$ is the mean virtuality of the exchanged $t$-channel gluon in the two-gluon exchange model.

\subsection{Nuclear targets}
For nuclear targets we compute the coherent cross section, in which the nucleus remains intact, by averaging over nucleon configurations in the optical approximation,
\begin{equation}
    \left\langle\frac{\dint\sigma_A}{\dint^2 b}\right\rangle
    =2\left[1-\left(1-\frac{T_A(b)}{2}\sigma_p\right)^{\!A}\right],
    \label{eq:optical}
\end{equation}
with $T_A(b)$ the transverse projection of the Woods--Saxon distribution and $\sigma_p$ the integrated proton dipole cross section. 

\section{Implementation in \sartre}
The complete coherent cross section is
\begin{equation}
    \frac{\dint^4\sigma^{ep/A}_{T,L}}{\dint Q^2\,\dint W^2\,\dint\beta\,\dint z}
    =\frac{\dint^2n_{T,L}^\gamma}{\dint W^2\dint Q^2}\,
    \frac{\dint^2\sigma_{T,L}^{\gamma^*p/A}}{\dint\beta\,\dint z},
\end{equation}
containing three separate contributions, $\sigma^{ep/A}_{T,q\bar q}$, $\sigma^{ep/A}_{L,q\bar q}$ and $\sigma^{ep/A}_{T,q\bar qg}$, with the difference that in the $q\bar qg$ case we generate $\tilde z$ rather than $z$. In \sartre\ these are interpreted as four-dimensional probability densities and used to generate $Q^2$, $W^2$, $\beta$ and $z$.

Once these kinematics and the beam four-momenta are fixed, the partonic final state is fully defined.  We treat the diffractive system $X$ as a pseudo-particle of mass $M_X$ and four-momentum $P_X=q+I\!\!P$, where $I\!\!P$ is the pomeron momentum with $I\!\!P^+=\xpom P^+$.  We then decay $X$ in its rest frame into a back-to-back quark--antiquark pair, uniformly in solid angle, and boost back to the lab frame.  For the $q\bar qg$ Fock state the procedure is repeated twice: $X$ is first decayed into a gluon and a pseudo-gluon of mass $M_{q\bar q}^2=Q^2(z/\beta-1)$, and the pseudo-gluon is then decayed into a $q\bar q$ pair.  The resulting partons are passed to \pythia\ for hadronisation with the Lund string model.

We do not currently calculate the cross section differentially in $t$, instead $t$ is generated in the final state from a form factor.  For nuclear targets we use the Woods--Saxon form factor
\begin{equation}
    F^{(A)}=\frac{4\pi\rho_0}{A\Delta^3}
    \big[\sin(\Delta R_A)-\Delta R_A\cos(\Delta R_A)\big]\frac{1}{1+a^2\Delta^2},
\end{equation}
where $R_A$ and $\rho_0$ are the Woods--Saxon radius and central density, $\Delta=\sqrt{-t}$ and $a=0.7$~fm.  For proton targets we use a Gaussian form factor $F^{(p)}\propto e^{-B_p\Delta^2/2}$.

\subsection{Cross-section lookup tables and machine learning}
To generate events efficiently we pre-compute four-dimensional lookup tables which serve as the probability density for event generation.  Since the final state carries information about which quark flavour participated, separate tables are created per flavour; using a common light-quark mass makes the down and strange tables identical, yielding nine 4D tables for four flavours.

We apply neural networks as described in~\cite{Singh:2023yvj} to allow the tables to be filled only partially and the remaining bins to be predicted from a model trained on the partial data. Since the tables contain the results of nested integrations, isolated kinematic points can exhibit numerical artefacts. The neural network model smooth the resulting  spikes, which would otherwise produce unphysical cross sections.

\section{Results}
\subsection{Comparison with HERA data}

\begin{figure}[htbp]
    \centering
    \includegraphics[width=0.8\linewidth]{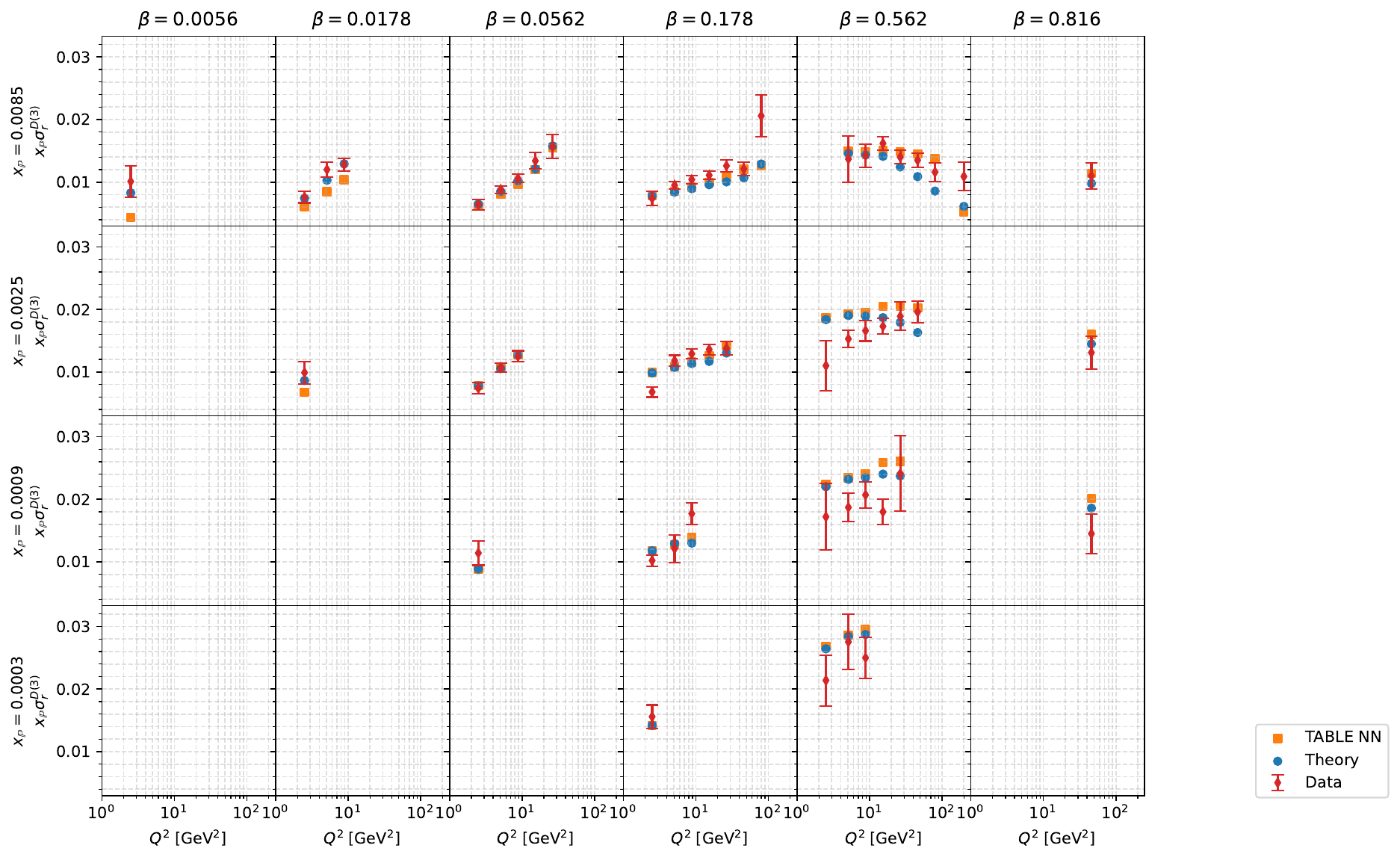}
    \caption{Comparison between the tabulated cross sections, smoothed with a
    neural network, and the combined H1 and ZEUS diffractive reduced cross
    section~\cite{H1ZEUS:2012diff}.}
    \label{fig:tables}
\end{figure}

Figure~\ref{fig:tables} compares the model of Sec.~\ref{sec:dipolemodel}, as tabulated for event generation and smoothed with a neural network, with the combined H1 and ZEUS measurement of the diffractive reduced cross section~\cite{H1ZEUS:2012diff}.  This is the first confrontation of the IPsat model with these data, and the model describes the measurements well, further supporting the universality of the dipole framework.

\begin{figure}[htbp]
    \centering
    \includegraphics[width=0.48\linewidth]{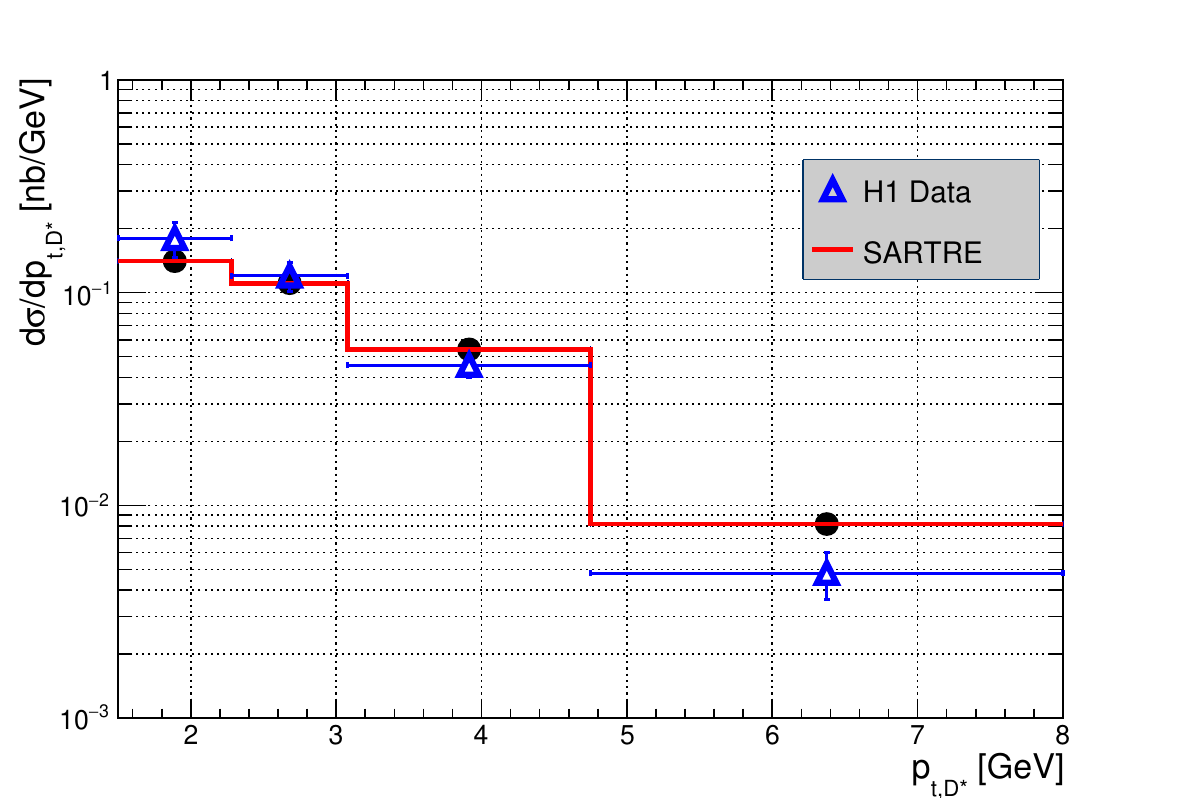}
    \includegraphics[width=0.48\linewidth]{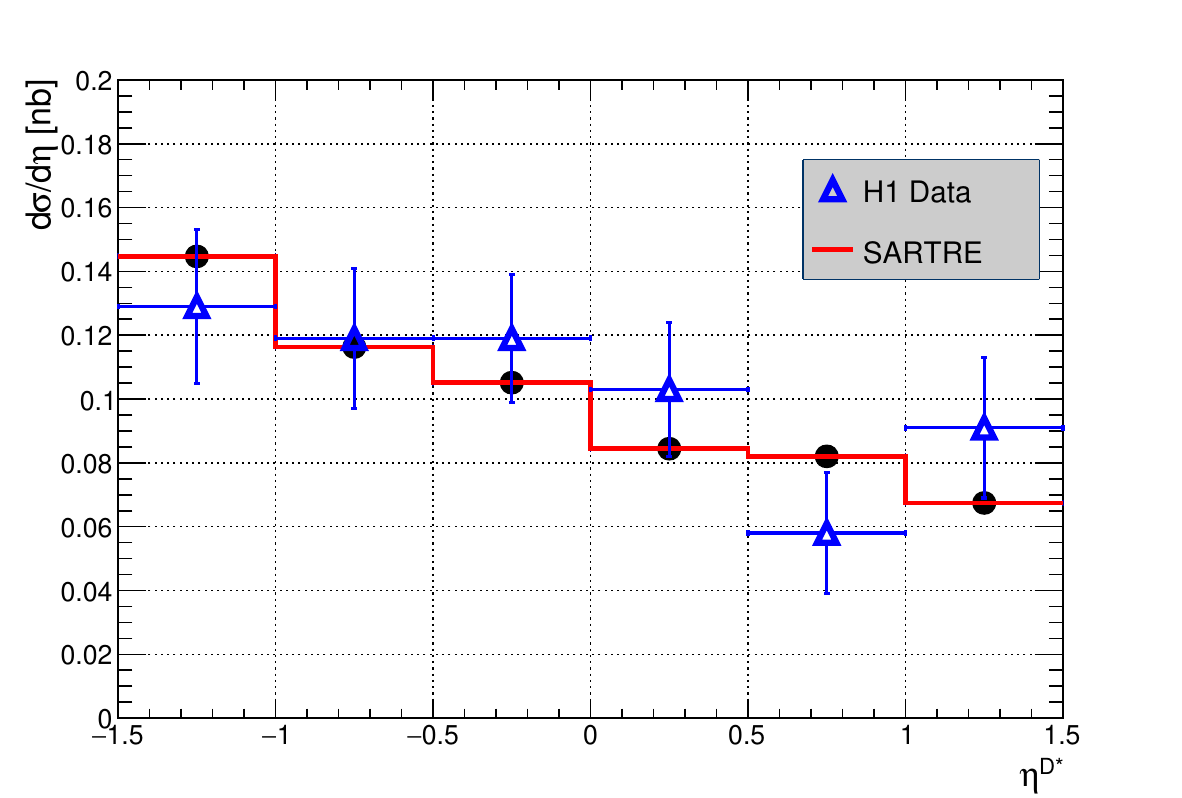}
    \caption{Comparison with diffractive $D^*$ production at H1~\cite{H1:2017bnb}
    in transverse momentum (left) and pseudorapidity (right).  The H1 data have
    been scaled by 0.3 to account for the $\xpom$ range mismatch described in
    the text.}
    \label{fig:Dstar}
\end{figure}

To validate the exclusive hadronic final state we compare with diffractive $D^*$ production at H1~\cite{H1:2017bnb} in Fig.~\ref{fig:Dstar}. H1 used data in the range $0.001\leq\xpom\leq0.03$, whereas the dipole model is reliable only for $\xpom<0.01$.  To make the comparison we define a $k$-factor by extrapolating the \sartre\ result into $0.01<\xpom<0.03$ and dividing by the H1 cross section over the same interval. This yields a factor of 0.3 by which the H1 data are scaled in Fig.~\ref{fig:Dstar}.  \sartre\ agrees well with the measured shapes in both transverse momentum and pseudorapidity, validating the final-state generation and hadronisation.

\subsection{Predictions for the Electron-Ion Collider}
\begin{figure}
    \centering
    \includegraphics[width=0.48\linewidth]{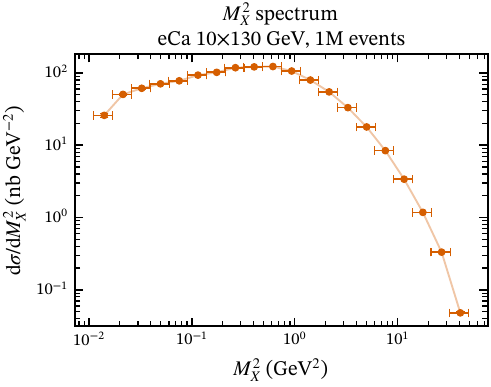}
    \includegraphics[width=0.48\linewidth]{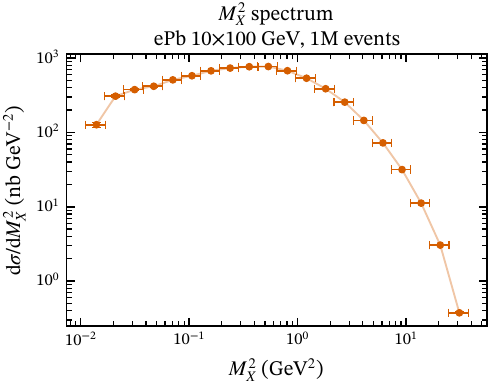}
    \caption{$M_X^2$ spectra from 1M events generated by \sartre\ in $e$Ca (left) and $e$Pb (right) collisions.}
    \label{fig:EICevents}
\end{figure}

We have generated lookup tables for $eA$ at EIC energies for $A=p$, $^{40}$Ca, $^{96}$Ru, $^{108}$Ag, $^{197}$Au and $^{208}$Pb. In Fig.~\ref{fig:EICevents} we show as examples the $M_X^2$ spectra from 1 million the $e$Ca and $e$Pb events at $10\times 130~{\rm GeV}$ and $10\times 100~{\rm GeV}$ respectively.

\section{Summary and Outlook}
We have presented our implementation of inclusive diffractive DIS for proton and nuclear targets at small $x$ into the \sartre\ event generator, using the IPsat and IPnonsat dipole model.  The model describes the combined HERA inclusive diffractive data, the first such comparison for IPsat, and the measured shapes of diffractive $D^*$ production. Since the diffractive cross section is roughly quadratic in the gluon density, and the IPsat/IPnonsat comparison isolates non-linear effects directly, inclusive diffraction in $eA$ is among the most sensitive saturation observables at the EIC, and \sartre\ enables realistic, fully exclusive simulations of it.  For future work we plan to include the incoherent nuclear cross section and as well as a complete calculation differential in $t$.

\acknowledgments
The authors acknowledge the support of the physics department of IIT Delhi.
This research was supported by Core Research Grant CRG/2022/002507 from the
Anusandhan National Research Foundation (ANRF), Department of Science and
Technology, Government of India.

\end{document}